\documentclass[conference]{IEEEtran}
\IEEEoverridecommandlockouts
\usepackage{cite}
\usepackage{amsmath,amssymb,amsfonts}
\usepackage{algorithmic}
\usepackage{graphicx}
\usepackage{textcomp}
\usepackage{xcolor}
\usepackage{epsfig,endnotes}
\usepackage{indentfirst}
\usepackage[font=small,labelfont=bf]{caption}
\usepackage{tabularx}
\usepackage{hyperref}
\def\BibTeX{{\rm B\kern-.05em{\sc i\kern-.025em b}\kern-.08em
    T\kern-.1667em\lower.7ex\hbox{E}\kern-.125emX}}
\begin{document}

\title{Automated labeling of bugs and tickets using attention-based mechanisms in recurrent neural networks}

\author{\IEEEauthorblockN{Volodymyr Lyubinets}
\IEEEauthorblockA{\textit{Forethought Technologies} \\
San Francisco, USA \\
vlyubin@gmail.com}
\and
\IEEEauthorblockN{Taras Boiko}
\IEEEauthorblockA{\textit{Lviv National University} \\
Lviv, Ukraine \\
me@tboiko.com}
\and
\IEEEauthorblockN{Deon Nicholas}
\IEEEauthorblockA{\textit{Forethought Technologies} \\
San Francisco, USA \\
deon@forethought.ai}
}

\IEEEspecialpapernotice{(Accepted to 2018 IEEE Second International Conference on Data Stream Mining and Processing)}

\maketitle

\begin{abstract}
We explore solutions for automated labeling of content in bug trackers and customer support systems. In order to do that, we classify content in terms of several criteria, such as priority or product area. 

In the first part of the paper, we provide an overview of existing methods used for text classification. These methods fall into two categories - the ones that rely on neural networks and the ones that don't. We evaluate results of several solutions of both kinds.

In the second part of the paper we present our own recurrent neural network solution based on hierarchical attention paradigm. It consists of several Hierarchical Attention network blocks with varying Gated Recurrent Unit cell sizes and a complementary shallow network that goes alongside.

Lastly, we evaluate above-mentioned methods when predicting fields from two datasets - Arch Linux bug tracker and Chromium bug tracker.

Our contributions include a comprehensive benchmark between a variety of methods on relevant datasets; a novel solution that outperforms previous generation methods; and two new datasets that are made public for further research.
\end{abstract}


\section{Introduction}

When dealing with a customer support ticket, one of the first things a customer service agent has to do is to label the ticket in terms of multiple criteria. These could be priority, product area, or whether action is required from an engineering team. Such labels are used for effective handling of the ticket - for example, tickets with high priority will be dealt with before low priority tickets, or engineering team will intervene only if the ticket was marked for intervention. Figure 1 shows an example of a labeling panel in one of these customer service platforms.

\begin{figure}[h]
    \centering
    \includegraphics[width=0.45\textwidth]{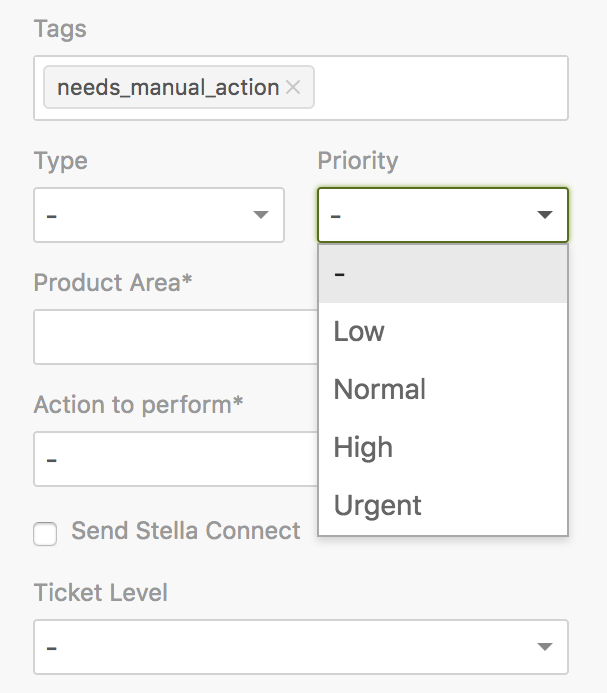}
    \caption{Setting labels for a customer service ticket}
    \label{fig:mesh1}
\end{figure}

A similar scenario is also true for project tracking systems such as JIRA or Bugzilla. Employees often label tasks in terms of area (e.g. kernel vs front-end) so that the appropriate team takes a look at it, or in terms of type - bugs typically have higher priority than new feature requests.

Considering that before such labeling takes place, hours or even days can pass, an ability to perform it automatically would increase the speed at which businesses operate and dramatically reduce the costs. Therefore, this is an extremely important problem, which got more traction recently due to evolution of deep neural networks and results achieved by leveraging word embeddings.

The general problem we are solving is that of text classification. Given a body of text we have to derive its class from a known fixed set of classes. However, using text classification on data from customer service systems and task trackers has its caveats. For example, such data often includes unique fragments that are hard for automatic systems to reason about, such as stack traces or HTML snippets. On the other hand, they often are well-structured and this structure can be leveraged by some of the novel methods, including the one we are proposing.

We are primarily interested in multi-class text classification, where number of classes we're predicting is larger than two. Historically, great results have been reached on binary classification tasks such as sentiment analysis (e.g. Twitter dataset, where you have to tell whether a tweet is positive or not), or spam filtering, where you have to tell whether email is a spam or no. Depending on the task, 95\%+ accuracies can often be achieved. This is due to fact that binary categories often have a lot of clue words present, which simplify classification task (e.g. "good" or "great" in positive reviews, or "sale" in spam email). However, if you look at data from Table 1, which shows state of the art results on multi-class classification dataset, it becomes clear that this is not a solved problem. Accuracies on Amazon and Yelp reviews datasets, where you have to predict the ranking of a review on 1 to 5 scale, hover around 60\% to 70\%, with nobody beating the 50\% threshold on IMDB dataset for movie genre prediction with 15 categories.

\bigskip
\begin{tabular}{ |p{2cm}||p{1.1cm}|p{1.1cm}|p{1.1cm}|  }
 \hline
 Paper& Yelp'15 & IMDB & Amazon\\
 \hline
 Zhang et al., 2015 &   59.9\%  & -   & 55.3\%\\
 Tang et al., 2015 &67.6\% & 45.3\% &  -\\
 Yang et al., 2016    &71.0\% & 49.4\%&  63.6\%\\
 \hline
\end{tabular}
\begin{center}
\textbf{Table 1:} State of the art accuracies on multi-class classification, results taken from \cite{YangYDHSH16}.
\end{center}

We being with an overview of existing methods, then we will present our own solution, and lastly we will provide a comprehensive benchmark of all methods in question.

\section{Existing methods}

Text classification is one of the most important problems of Natural Language Programming (NLP) research and variety of methods have been proposed for it.

These methods can be split into two large families - classic solutions that don't leverage neural networks, and novel solutions that leverage recurrent neural networks, especially with the use of word embeddings. Among the former methods are Naive Bayes and algorithms that use term count data, such as Term Frequency Inverse Document Frequency (TF-IDF) fed into the Support Vector Machine (SVM) classifier. Among the latter are methods that use recurrent neural networks on word embeddings data, which differ in network structure, loss functions, algorithms used to derive embeddings and preprocessing routines.

In this section we provide a brief overview of these methods, as they will be part of the benchmark in the section 4.

\subsection{Naive Bayes}

Naive Bayes used to be one of the most popular algorithms for text classification, coming into NLP scene in 1960's. It was widely used for early spam filters, where it still performs fairly well \cite{Metsis2006SpamFW}. But as we will see later, Naive Bayes shows poor results on multi-class classification.
The key idea behind Navie Bayes classifier is using the Bayes theorem - for a document \textit{d} and class \textit{c}, we can say that probability of that document having class \textit{c} is: \[ P(c|d) = \frac{P(d|c)P(c)}{P(d)} \]
After making the "naive" assumption about independence of conditional probabilities for individual terms, we get \[ P(d|c) = P(x_1, x_2, ..., x_k|c) = P(x_1|c)P(x_2|c) ... P(x_k|c) \], where \(x_i\) are terms contained in document \(d\). Then we simply choose the class that maximizes \(P(x_1, x_2, ..., x_k|c)P(c)\), with each of these probabilities computed on the training set.

\subsection{TF-IDF with SVM}

Since 1990's algorithms using term count statistics such as TF-IDF took prominence in NLP community. The idea behind TF-IDF is to represent each sentence as a vector of scores determined by term frequencies. The score consists of two parts - one determined by counts of that term inside a document and the other by presence of the term across the body of documents, with the score being the multiplication of the two .

Once we have the TF-IDF data, we can use it with any supervised classification method, such as Softmax classifier or Support Vector Machine. The latter is a popular choice and used to be state of the art method before emergence of neural networks \cite{Pilszy2005TextCA}. We evaluate SVM with linear kernel on our data in section 4, where it shows good results.

We have also tried TF-IDF data with several other classifiers, such as neural networks, but found the results to be worse than that with Support Vector Machines. This is due to the fact that neural networks overfit quite easily on sparse data like TF-IDF, while SVMs are unable to achieve perfect fit on it and thus act as a "natural" regularizer.

\subsection{Word embeddings and Recurrent Neural Networks}

With the introduction of Mikolov et al. \cite{MikolovSCCD13} paper in 2013, the vector of NLP research turned towards word embeddings. The idea behind word embeddings is to represent each word with a vector, rather than the entire document as TF-IDF does. The general idea behind how these embeddings are computed is that words that occur together a lot should have similar values (as determined by an appropriate loss function), while those that rarely occur together should be different. Refer to \cite{mikolov2013efficient} and \cite{MikolovSCCD13} for more details about the training process. Word embeddings are an excellent candidate to be used with recurrent neural networks (RNN), with each embedding vector typically used as one of the inputs to the first RNN in the stack.

One of the libraries that provides an efficient way to compute word embeddings is fastText \cite{joulin2016bag}. In addition to that, it provides an out of the box classification solution, which we are going to evaluate on our datasets.

The architecture of classification fastText is a vanilla many-to-one RNN that takes word embeddings as inputs, and the resulting output fed into a Softmax classifier.

\subsection{Solution by DeepTriage}

Another solution using recurrent networks and word embeddings, that was built to perform bug triaging is DeepTriage from \cite{deeptriage}. Considering a similarity of their use case (triaging can be considered as an extreme multi-class classification, with number of classes being in the hundreds), it is an excellent candidate to benchmark against.

The architecture of DeepTriage consists of a bidirectional RNN, followed by two fully-connected layers. While in the paper they mention using soft attention modules, the provided code does not use them by default. DeepTriage is going to be used as another candidate for our benchmark.

\section{Proposed approach}

The solution that we propose is based on using hierarchical attention paradigm with varying Gated Recurrent Unit (GRU) \cite{cho2014learning} cell sizes, and a shallow network that goes alongside. This allows network to outperform regular hierarchical attention on datasets where simpler term-based approaches work well.

\subsection{Preprocessing pipeline}

Before we begin the overview of our solution, it's worth mentioning the preprocessing that we have done with the data. This preprocessing routine was shared across all approaches in the benchmark, as each of them has shown better results on the preprocessed data. In general, data cleaning is an extremely important step when developing a machine learning solution, and this is especially true for data in customer service systems and task trackers. For example, in the two datasets that we will be using for benchmarking, people often include stacktraces and error messages. And while error messages carry some weight, stacktraces in most cases are meaningless numbers that only add noise. At Forethought Technologies, we have seen similar issues with customer support tickets, which often contain HTML snippets.

Our final pipeline includes the following steps:

\begin{itemize}
  \item Casting everything to lowercase. 
  \item Removing stopwords.
  \item Filtering dataset-specific garbage. This was done by custom regex expressions created upon inspection of the datasets.
\end{itemize}

We have tried several other common preprocessing routines such as stemming, but they have not led to improved results.

\subsection{Hierarchical Attention}

Hierarchical Attention is an approach proposed by Yang et al in \cite{YangYDHSH16} and it consists of two key ideas - use of sentence hierarchy, and use of attention vector.

The idea of using sentence hierarchy means that we are going to use one RNN that takes in word embeddings from a particular sentence as inputs and compute another vector that acts as a representation of that sentence. Afterwards, a second RNN will take those sentence vectors, and compute a final vector for the document, that will be passed into the Softmax layer to derive final probabilities. Considering that language is structured in sentences, this paradigm works quite nicely in practice, with authors able to beat the best result on Yahoo Q\&A dataset with using this approach alone (and without the use of attention vectors). Figure 2 contains an architecture diagram from the paper.

Considering that documents in a dataset can often follow a particular structure (e.g. the most important information is located in the end), it would be good to have appropriate coefficients for outputs of both word and sentence encoders. This task is done by introducing attention vectors, which are marked as \(u_s\) and \(u_w\) in the diagram. They are shared across all outputs at their level and are trained alongside other parts of the model. When it comes to combining sentence or word vectors into one, the coefficient that we are going to use will be a dot product of an appropriate attention vector with sentence or word. So attention vector serves as "the ideal vector", which if present would achieve perfect score. Using attention vectors in our networks makes a lot of sense, as data is often well structured, with a significant portion of items in the Linux Bugs Dataset (see section 4) filling out a predefined template for their bugreport. Such scenario is perfect to be used with attention-based mechanisms.

It is worth mentioning that idea of using hierarchy for detection and classification is not new and has been successfully applied in other fields, such as visual object recognition \cite{haother}.

\begin{figure}[h]
    \centering
    \includegraphics[width=0.45\textwidth]{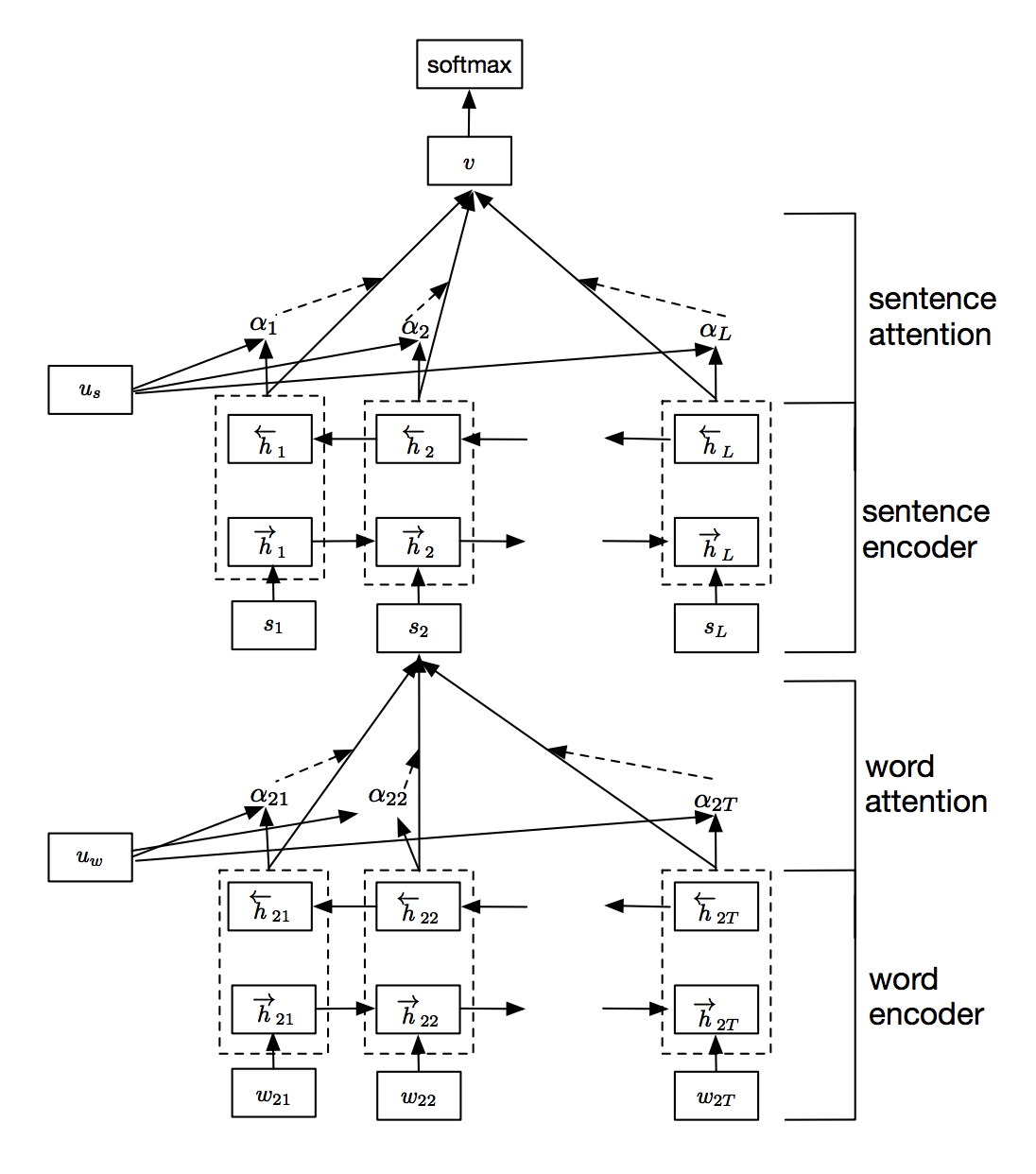}
    \caption{Architecture of network in Hierarchical Attention paper}
    \label{fig:mesh1}
\end{figure}

\subsection{Network Architecture}

As you will see in the benchmark, hierarchical attention does well on a dataset that contains well structured data (Linux bugs), but performs poorly on a dataset where such structure doesn't exist. To combat these problem, we introduce two changes. 

First, we are going to use several hierarchical attention blocks like the ones you see on Figure 2, with each of them having a different GRU cell size. Hierarchical attention uses GRU cells rather than more traditional Long Short Term Memory cells (LSTM) \cite{lstm}, claiming higher performance, albeit with a small margin. The architecture of one such block is depicted on Figure 3. These Deep Attention Blocks will be used for both word-level and sentence-level processing afterwards.

Second, we are going to introduce an additional "shallow" network that is a just a simple RNN that takes in word embeddings and produces one vector. It is using GRU cells as well.

Afterwards, the outputs from the shallow network and deep attention blocks are stacked together and go into the fully connected layer, and then into the Softmax layer that produces the final result.

We use cross-entropy loss function for training:

$$L{y'} (y) := - \sum_{i} y_{i}' \log (y_i)$$

where $y_i$ is the predicted probability value for class $i$ and $y_i'$ is the true probability for that class.

Our network structure is presented on Figure 4.

\begin{figure}[h]
    \centering
    \includegraphics[width=0.35\textwidth]{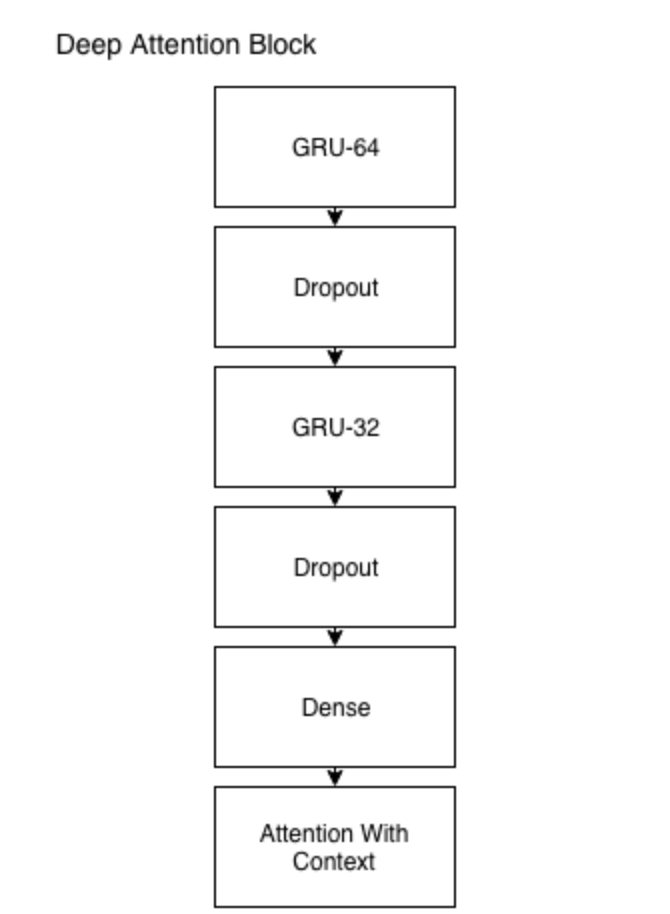}
    \caption{Deep Attention Block}
    \label{fig:mesh1}
\end{figure}

\begin{figure}[h]
    \centering
    \includegraphics[width=0.35\textwidth]{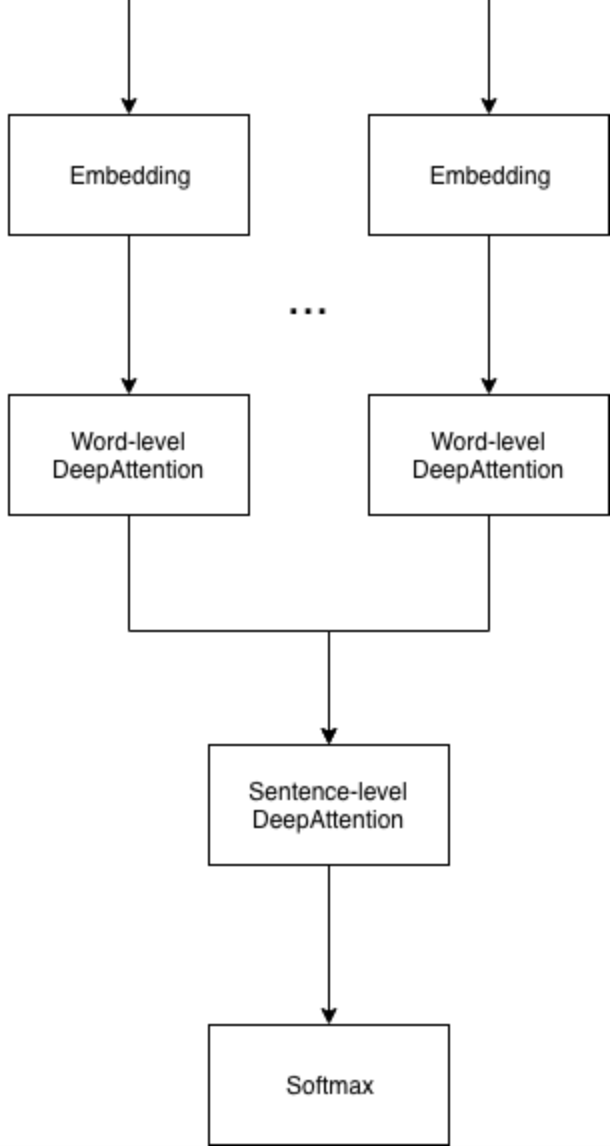}
    \caption{Our Network Architecture (without an auxiliary shallow network)}
    \label{fig:mesh1}
\end{figure}

\subsection{Training Details}

Training deep neural networks can often be a finicky task, so we would like to mention several details from training our solution. First, we extensively use dropout \cite{srivastava2014dropout} to avoid overfitting, which could easily happen considering datasets sizes. What is even more interesting is that we found dropout probability to work best when set at around $1/2$, which is higher than typical values. Dropout layers are present in between any two RNN or affine layers in our solution (e.g. see Figure 3). Second, we use RMSprop \cite{Tieleman2012} method for optimization. Lastly, word embeddings that we use for our RNNs are computed by Word2vec.

\section{Comparative benchmark}

Next, we would like to explore how the aforementioned solutions, including our own, perform on the real data. For this purpose, we have collected two datasets, that are made public for further research. They and the code behind solutions in the benchmark is available at at \url{https://github.com/Forethought-Technologies/ieee-dsmp-2018-paper}.

\subsection{Arch Linux bugtracker dataset}

The first dataset that we are going to use contains bugs from open Arch Linux bugtracker at \url{https://bugs.archlinux.org}. We wrote a simple web scraper to acquire this data. Just like with other bug tracking systems, authors label bugs in terms of various criteria, and two such criteria that we are going to predict are priority and product area. It is easy to see a practical use case for a system that can predict such fields, as the former allows to establish priority of which bugs to fix first and the latter allows to pinpoint the team best suited for the task. Priority field has 9 classes (from P1 high to P3 low), the product field has 16 classes (Network, Drivers, etc.). Below is an example of a bug from this dataset:

\smallskip
\textbf{Title}: \textit{i2o\_scsi does not handle reset properly}

\textbf{Content}: \textit{The i2o scsi driver should sleep in the reset handler until the i2o reset message is replied to by the firmware. James has discussed infrastructure to make this generic}

\textbf{Priority}: P2\_low (9 classes)

\textbf{Product}: Drivers (16 classes)
\smallskip

The complete dataset contains 16,456 entries.

\subsection{Chromium bugtracker dataset}

The second dataset was adapted from the Chromium dataset used by DeepTriage paper. The field that we are going to predict is called "Type" and can be one of Bug / Feature / Compatibility issue.

\smallskip
\textbf{Title}: \textit{Scrolling with middle-mouse button does not work (autoscroll)}

\textbf{Content}: \textit{Product Version: chrome beta 1
URLs (if applicable) :
Other browsers tested:
Add OK or FAIL after other browsers where you have tested this issue: Safari 3: OK Firefox 3: OK IE 7: OK
What steps will reproduce the problem?
What is the expected result? Clicking the middle-button on the mouse should show a ""fast scroll"" feature.
What happens instead? Nothing.
Please provide any additional information below. Attach a screenshot if possible.}

\textbf{Type}:  Feature
\smallskip

The complete dataset size contains 58,871 entry.

\subsection{Evaluation methodology}

For each solution we measure two results - accuracy and weighted F1 score. The results are computed on the test set, with test set size being 15\% of the original data.

Each classifier had optimal hyperparameters picked via a usual grid search.

\smallskip
\begin{tabularx}{\textwidth}{X}
\begin{tabular}{ |p{4.9cm}||p{3.2cm}|p{3.2cm}|p{3.2cm}|  }
 \hline
 Method& Linux bugs: Importance (9 classes) & Linux bugs: Product (16 classes) & Chromium bugs: Type (3 classes)\\
 \hline
 Naive Bayes\endnote{Naive Bayes was ran on subset of the data due to slow performance} &   51.6\%  & 45.6\%   & 80.5\%\\
 TF-IDF with SVM &65.0\% & 61.6\% &  80.5\%\\
 fastText    &64.2\% & 58.7\%&  82.2\%\\
 DeepTriage\endnote{This and following solutions need a GPU machine to run. We have used p2.xlarge on Amazon AWS, which has Tesla K80 GPU.}    &61.4\% & 63.8\%&  81.6\%\\
 Hierarchical Attention (regular)    &66.4\% & 58.9\%&  75.9\%\\
 Our Solution    &69.1\% & 58.7\%&  88.2\%\\
 \hline
\end{tabular}
\begin{center}
\textbf{Table 2:} Benchmark of accuracies.
\end{center}
\end{tabularx}

\smallskip
\begin{tabularx}{\textwidth}{X}
\begin{tabular}{ |p{4.9cm}||p{3.2cm}|p{3.2cm}|p{3.2cm}|  }
 \hline
 Method& Linux bugs: Importance (9 classes) & Linux bugs: Product (16 classes) & Chromium bugs: Type (3 classes)\\
 \hline
 Naive Bayes &   0.479  & 0.411   & 0.787\\
 TF-IDF with SVM &0.568 & 0.590 &  0.804\\
 fastText    &0.542 & 0.579&  0.821\\
 DeepTriage    &0.516 & 0.604&  0.816\\
 Hierarchical Attention (regular)    &0.573 & 0.574&  0.758\\
 Our Solution    &0.579 & 0.567&  0.879\\
 \hline
\end{tabular}
\begin{center}
\textbf{Table 3:} Benchmark of F1 scores.
\end{center}
\end{tabularx}

\subsection{Results}

Results are presented in tables 2 and 3. We can make several conclusions from these results:
\begin{itemize}
  \item Naive Bayes is not a good solution for multi-class classfication.
  \item Decade old solutions like TF-IDF with SVM still show pretty good results and can be a great solution for cases where resources are limited.
  \item Novel solutions can outperform the classic ones on each dataset.
  \item Our solution shows superior performance, especially on the last task. However, it did poorly when predicting the product field from the first dataset.
\end{itemize}

\section{Conclusion}

In this paper we went over a variety of methods used for text classification, presented our own solution based on hierarchical attention paradigm and benchmarked these solutions on two real datasets. We see that novel approaches that use RNNs on word embeddings data outperform the classic solutions, which opens doors to many practical use cases. Nevertheless, the resulting accuracies are still far from perfect and multi-class text classification remains an open problem.

\bigskip
\bigskip
\bigskip
\bigskip
\bigskip
\bigskip
\bigskip
\bigskip
\bigskip
\bigskip
\bigskip
\bigskip
\bigskip
\bigskip
\bigskip
\bigskip
\bigskip
\bigskip
\bigskip
\bigskip
\bigskip
\bigskip
\bigskip
\bigskip
\bigskip
\bigskip
\bigskip
\bigskip
\bigskip
\bigskip
\bigskip
\bigskip
\bigskip
\bigskip
\bigskip
\bigskip
\bigskip
\bigskip
\bigskip
\bigskip

{\footnotesize \bibliographystyle{acm}
\bibliography{bib}}

\theendnotes

\end{document}